\documentclass[onecolumn]{aa} 
\usepackage{txfonts}
\usepackage{graphicx}
\def\apj {ApJ}
\def\aa {A\&A}
\def\apjs {ApJs}
\def\mnras {MNRAS}
\def\aj {AJ}
\begin{document}
\title{GMRT and VLA observations of \ion{H}{i} and OH from the Seyfert galaxy Mrk~1}
\author{A. Omar
          \inst{1}\thanks{aomar@rri.res.in}
          \and
	  K.S. Dwarakanath\inst{1,2}\thanks{dwaraka@rri.res.in}
	  \and
	  M. Rupen\inst{2}\thanks{mrupen@aoc.nrao.edu}
	  \and
	  K.R. Anantharamaiah\inst{1}\thanks{Deceased}
          }

\offprints{A. Omar}
\institute{Raman Research Institute, C.V. Raman Avenue, Bangalore, 560 080, India. 
\and
National Radio Astronomy Observatory, P.O. Box O, Socorro, NM 87801, USA \\
		}
\authorrunning{A. Omar et al.}
\titlerunning{Mrk~1 -- GMRT and VLA observations}
\date{Received 19 June 2002 / Accepted 02 August 2002}

\abstract   {We  present  Giant   Meterwave  Radio   Telescope  (GMRT)
observations of  the \ion{H}{i}~21~cm line and Very  Large Array (VLA)
observations  of the OH~18~cm  line from  the Seyfert~2  galaxy Mrk~1.
\ion{H}{i}  emission is  detected from  both Mrk~1  and  its companion
NGC~451. The \ion{H}{i} emission  morphology and the velocity field of
Mrk~1  are disturbed.   We speculate  that the  nuclear  activities of
Mrk~1 are triggered by tidal interactions.  We estimate the \ion{H}{i}
masses of Mrk~1  and NGC~451 to be $8.0(\pm0.6)\times10^{8}$~M$_\odot$
and  $1.3(\pm0.1)\times10^{9}$~M$_\odot$ respectively.   We  have also
detected the \ion{H}{i}~21~cm line and the OH~18~cm line in absorption
toward the nucleus of Mrk~1  at a blueshifted velocity with respect to
its systemic  velocity indicating an  outflow of atomic  and molecular
gas. Two  OH lines, at 1665  and 1667~MHz, are detected.   Each of the
profiles  of  the  \ion{H}{i}   and  OH  absorption  consists  of  two
components  that are  separated  by $\sim$~125~km~s$^{-1}$.   Gaussian
fitting  gave   dispersions  of  $\sim44$~km~s$^{-1}$   for  both  the
components  of  the \ion{H}{i}  absorption.   The  profile  of the  OH
absorption  is  qualitatively  similar   to  that  of  the  \ion{H}{i}
absorption.   Both  components  of  the OH  absorption  are  thermally
excited.   The  peak optical  depths  of  the  two components  of  the
\ion{H}{i}     absorption    are     $(7.3\pm0.4)\times10^{-2}$    and
$(3.2\pm0.4)\times10^{-2}$.  The corresponding  peak optical depths of
the   1667~MHz  OH   absorption  are   $(2.3\pm0.3)\times10^{-2}$  and
$(1.1\pm0.3)\times10^{-2}$.   The higher  velocity  components of  the
\ion{H}{i} and OH (1667~MHz) absorption lines are blueshifted from the
[\ion{O}{iii}]$\lambda5007$,    [\ion{O}{i}]$\lambda6300$,   and   the
systemic velocity  by $\sim$~100~km~s$^{-1}$, but  are consistent with
the  [\ion{O}{ii}]$\lambda3727$ velocity.   We explain  these velocity
discrepancies as due  to shock ionization of a  region which is pushed
forward due  to shocks  in front of  the radio nucleus  thereby giving
apparent    blueshift   to    \ion{H}{i},   OH,    and   [\ion{O}{ii}]
velocities.          The         optical          depth         ratios
$\tau_\mathrm{\ion{H}{i}}/\tau_\mathrm{OH}^{1667}$    of    both   the
components  of   the  \ion{H}{i}  and  OH   absorption  are  $\sim$~3,
indicating   their   origin   in   dense  molecular   clouds.    Using
OH/A$_\mathrm{v}$ values for the  Galactic molecular clouds, we obtain
9  $<$  A$_\mathrm{v}<$  90  toward   the  line  of  sight  of  Mrk~1.
 } 
\maketitle

\section{Introduction} 

 Both the AGN  and the nuclear starburst activities  in galaxies which
require  inflow of  material  toward  the centre  either  to fuel  the
central black hole  or to cause rapid burst  of nuclear star formation
can be  accomplished by tidal interactions (\cite{her95}).   It is not
clear,  however,  in the  case  of  Seyfert  galaxies whether  nuclear
activities in  these low luminosity  active galactic nuclei  (AGN) are
due to interactions as found in QSOs, radio galaxies, and BL Lacs (see
\cite{der98} for  a review on  the subject). It is  generally accepted
that interactions  leading to mergers (bound interactions)  may play a
significant  role in  triggering  nuclear activities  than unbound  or
hyperbolic encounters (\cite{der98}).  Interactions can be effectively
traced via \ion{H}{i} 21~cm  line emission from galaxies as \ion{H}{i}
disks often extend well beyond the optical radii of galaxies where the
disks  respond  quickly  to  gravitational  perturbations.  \ion{H}{i}
emission  studies may  be  particularly useful  since  most often  the
\ion{H}{i}  morphology provides  evidences of  interactions  which are
undetectable at optical wavelengths (e.g., \cite{sim87}).

\ion{H}{i}  in absorption  can  trace kinematics  and distribution  of
atomic gas near  the centres of active galaxies on  the size scales of
their background radio sources. The advantage of absorption studies is
that it can detect relatively  small quantities of gas irrespective of
the  redshift  of  the  object.   Recently,  \cite{gal99}  have  found
\ion{H}{i} rich absorbing disks on the scales of a few hundred parsecs
in  several  Seyfert  galaxies.    As  a  result  of  intense  nuclear
activities,  gas in  the central  regions  of active  galaxies may  be
perturbed due to interactions of the radio plasma with the surrounding
ISM which may result in bulk outflows of material (e.g., \cite{tad01},
\cite{mor98}). The  molecular gas near the centres  of active galaxies
can  be  traced  via  18~cm~OH  line  in absorption.   The  18  cm  OH
absorption line is sensitive to  molecular gas in both the diffuse ISM
and in the dark clouds with  OH to H$_{2}$ ratio being almost constant
over a  large variety of Galactic clouds  (\cite{lis96}). Studies have
shown that chances  of detecting OH absorption are  higher in infrared
luminous galaxies (Schmelz et al. 1986).

In   this   paper,   we   present  synthesis   observations   of   the
\ion{H}{i}~21~cm  line obtained with  the GMRT  and the  OH~18~cm line
obtained with the VLA of the infrared luminous active galaxy Mrk~1 and
its companion  NGC~451. The global properties of  Mrk~1 are summarized
in the next section. The details of observations and data analyses are
given  in  Sect.~3. The  results  are  presented  in Sect.~4.  Sect.~5
discusses  the radio  continuum properties,  \ion{H}{i}  emission, and
\ion{H}{i} and OH absorption. The conclusions are in the last section.

\section{Global properties of Mrk~1}

\begin{table*}
\begin{center}
\caption{\bf{Global properties of Mrk~1}}
\label{tab:mrk01}
\begin{tabular}{lcc}
\hline
\hline
\bf{Parameter} & \bf{Value} & \bf{Reference} \\
\hline
Right Ascension (J2000) & $01^\mathrm{h}16^\mathrm{m}07\fs1$ & 1\\
Declination (J2000) & 33\degr05\arcmin22\arcsec &1 \\
Distance (Mpc)  & 68 &2 \\ 
Hubble type & SB O/a & 1\\
Seyfert type & 2 &1\\
Inclination & 45\degr & 3 \\
Optical diameter (kpc) &$9.0\times5.3$ &1 \\
Corrected blue magnitude B$_\mathrm{T}^\mathrm{o}$ & 14.53 &4 \\
Total blue luminosity (L$_{\odot}$) &$1.1\times10^{10}$ & \\
Total \ion{H}{i} mass (M$_{\odot}$) &$8.0\times10^{8}$ & 5 \\
\ion{H}{i} mass to blue luminosity ratio (M$_{\odot}$/L$_{\odot}$) &0.07 &5 \\
Total FIR luminosity (L$_{\odot}$)&$1.7\times10^{10}$ &6 \\
1.4~GHz radio luminosity  (W Hz$^{-1}$) &$4.2\times10^{22}$&5 \\
Spectral index (S$~\propto~\nu^{-\alpha}$) (2.7 GHz -- 10.5 GHz) &0.8 &7 \\
X-ray luminosity (erg s$^{-1}$) &$<10^{41}$ & 8 \\
Systemic velocity (km~s$^{-1}$) &4780$\pm$2   &9 \\
$[\ion{O}{iii}]~\lambda$ 5007 velocity (km~s$^{-1}$) &4822$\pm$25  &10  \\
$[\ion{O}{ii}]~\lambda$ 3727 velocity (km~s$^{-1}$) &4697$\pm$25  &10 \\
$[\ion{O}{i}]~\lambda$ 6300 velocity (km~s$^{-1}$) &4817$\pm$25  &10 \\
Mean velocity of \ion{H}{i} emission (km~s$^{-1}$) &4780 &5 \\
H$_{2}$O maser velocity (km~s$^{-1}$) &$4868\pm1$ &3\\
Mean velocity of CO emission (km~s$^{-1}$) &4850 &11 \\
\hline \\
\multicolumn{3}{p{5in}}{Notes:     H$_{0}$     =    75     km~s$^{-1}$
Mpc$^{-1}$. The velocity definition is optical and Helio-centric.}\\
\multicolumn{3}{p{5in}}{1:  Markarian  et  al.  (1989);  2:  White  et
al.  (1999); 3  Braatz et  al.   (1997); 4:  NED (NASA  Extra-galactic
Database);  5:This  paper; 6:IRAS  faint  source  catalog, (1990);  7:
Dickinson et al.  (1976); 8: Fabbiano et al.   (1992); 9: Keel (1996);
10: De Robertis \& Shaw (1990) 11: Vila-Vilaro et al. (1998)}
\end{tabular} 
\end{center} 
\end{table*}

The     global     properties    of     Mrk~1     are    listed     in
Table~\ref{tab:mrk01}. Mrk~1 (NGC~449;  B$_\mathrm{T}^{0}$ = 14.53) is
a  member  of   a  poor  group  (WBL~035)  at   a  redshift  of  0.017
(\cite{whi99}).  The other two members of this group viz., NGC~447 and
NGC~451, are at projected  separations from Mrk~1 of $\sim$~38~kpc and
$\sim$~130~kpc respectively.  Mrk~1 is  classified as a  SB~O/a galaxy
with  a Seyfert  type 2  nucleus  (\cite{FBS}) with  no signatures  of
interactions in the optical images.  Mrk~1 is also a luminous Infrared
galaxy (L$_{\mathrm{FIR}}  = 1.7\times10^{10}$L$_{\odot}$), indicating
a high rate of star formation. Mrk~1 is one among 16 galaxies detected
in  the 22~GHz  water megamaser  emission in  a sample  of  354 active
galaxies (\cite{bra94}). The nuclear optical spectrum of Mrk~1 studied
by \cite{kos78}  and \cite{wee68} shows strong  emission lines typical
of an active  galaxy photo-ionized by hard continuum.  The broad lines
indicative of  a hidden  Seyfert nucleus are  not found either  in the
infrared  (\cite{vei97})  or in  the  polarized light  (\cite{kay94}).
High      dispersion     spectroscopic     observation      of     the
[\ion{O}{iii}]$\lambda5007$ line by \cite{ber87} shows a distinct blue
asymmetry indicative of an outflow of gas. \cite{kee96} suggested that
the nuclear activities of Mrk~1 are due to an ongoing interaction with
the nearby galaxy NGC~451.

The  radio continuum  emission from  Mrk~1 is  known to  have  a steep
spectrum with a spectral index $\alpha$~(S$~\propto~\nu^{-\alpha}$) of
0.8 (\cite{dick76}).  The  1.6~GHz EVN image (resolution $\sim$~30~pc)
of  \cite{kuk99}  shows  the   nuclear  emission  to  consists  of  an
unresolved core  surrounded by  a weak diffuse  emission with  a total
flux density of 34~mJy.  The NVSS flux density at 1.4~GHz is 75.4~mJy

The Arecibo observations  by \cite{hut89} detected \ion{H}{i} emission
and blueshifted  \ion{H}{i} absorption  from Mrk~1.  This  single dish
spectrum  could  not  separate  \ion{H}{i}  emission  from  Mrk~1  and
NGC~451.   Observations  with the  Nobeyama  Radio Telescope  detected
CO~(J   =   1--0)   emission   with   a   total   flux   integral   of
$11.5\pm1.6$~K~km~s$^{-1}$  from  the central  5~kpc  region of  Mrk~1
(\cite{vil98}).    The   search  for   the   18~cm~OH  absorption   by
\cite{sch86} with  the Arecibo  reflector resulted in  a non-detection
with an rms sensitivity to an optical depth of 0.02.

\section{Observations and data analyses}
\subsection{The GMRT observations}

\begin{table*}
\centering
\caption{\bf{Observational parameters}}
\label{tab:obspar}
\begin{tabular}{lrr}
\hline
\hline
\bf{Parameter} & \bf{GMRT}& \bf{VLA} \\
\hline
Dates of Observations &2000 Oct 25, 28 & 2001 Mar 26, 27 \\
Pointing centre (RA J2000.0) &$01^\mathrm{h}16^\mathrm{m}07\fs25$  
&$01^\mathrm{h}16^\mathrm{m}07\fs25$  \\
Pointing centre (Dec J2000.0) &+33\degr05\arcmin22\farcs2 
&+33\degr05\arcmin22\farcs2\\
Observing duration (hrs) &8 &4.5$^{\dagger}$ \\
Range of baselines (km) & 0.1--25& 0.1--11 \\
Observing frequency (MHz) &1395.00 &1640.22 \\
Bandwidth per IF (MHz) &8.0 &6.25 \\
Number of spectral channels &128 &128 \\
Polarizations &2 & 1\\
Frequency resolution (kHz) &62.5 & 48.8 \\
Velocity resolution (km~s$^{-1}$)&13.7 &9.0 \\
Amplitude calibrator &0137+331 &0137+331 \\
Phase calibrator &0137+331  & 0137+331  \\
Bandpass calibrator &0137+331  &0137+331  \\
\hline
$^{\dagger}$~Usable time, see section 3.2 for details & & \\
\end{tabular}
\end{table*}

The GMRT  observations of Mrk~1 were  carried out in  October, 2000. A
summary   of  the   main   observational  parameters   are  given   in
Table~\ref{tab:obspar}.  At the time of the observations, the GMRT was
not fully operational and hence not all 30 antennas were available for
observations at any given time.   Two runs of observations with 18--20
antennas,  each with  a field  of view  (FWHM) 24\arcmin  ~centered on
Mrk~1, were carried out on two  different days.  The GMRT has a mix of
both  short and long  baselines (see  \cite{swa91} for  more details),
making  it  sensitive  to  diffuse  emission  of  extent  as  much  as
$7\arcmin$  while having  a maximum  resolution of  $\sim~3\arcsec$ at
1.4~GHz.   The GMRT  uses a  30-station FX  correlator  which produces
complex visibilities  over 128  spectral channels in  each of  the two
polarizations.   The  bandwidth can  be  selected  in  multiples of  2
between 62.5~kHz and 16~MHz.  These observations were carried out with
a bandwidth of 8~MHz  centered at 1395.0~MHz, which covered \ion{H}{i}
velocities  in  the  range  $3730-5460$~km~s$^{-1}$  with  a  velocity
resolution of $\sim$~14~km~s$^{-1}$.
 
The  complex gains  of the  antennas were  determined every  30 minute
using  observations of an  unresolved nearby  ($\sim~4.5\degr$) source
(3C~48)  for 5  minutes. 3C~48  was  also used  for the  flux and  the
bandpass  calibrations.  The  data  were reduced,  following  standard
calibration  and   imaging  methods,  using   the  Astronomical  Image
Processing  Software (AIPS)  developed  by the  NRAO.   The data  were
calibrated for  the amplitude, phase,  and frequency response  for all
antennas separately for each  polarization.  The flux density of 3C~48
was estimated  to be  16.228 Jy at  the observing frequency  using the
1999.2 VLA flux densities of the standard VLA flux calibrators and the
formula given in the AIPS task  `SETJY'. Due to the close proximity of
3C~48 to Mrk~1 and based  on some previous test experiments, we expect
that the flux calibration is accurate to within 5\%.

A   continuum  data  set   was  formed   by  averaging   80  line-free
channels. The data  were self calibrated in both  phase and amplitude.
The   resulting  antenna   gain  corrections   were  applied   to  all
channels.  The continuum  images were  made using  the self-calibrated
averaged data from the line-free channels.  The continuum flux density
from each individual channel was  subtracted in the (u,v) dataset by a
linear fit to the visibilities  in the line-free channels. Since these
observations  were also  sensitive  to \ion{H}{i}  emission, the  data
points  were ``natural-weighted"  to enhance  sensitivity  to extended
features.   The  resulting spectral  cubes  were  CLEANed for  signals
greater than  4 times the rms  noise in the channel  images.  The cube
was blanked  for emission below a  level of 1.5$\sigma$  in the images
after  applying a  Hanning smoothing  of three  velocity  channels and
Gaussian  smoothing  of five  pixels  (pixel  size~=~6\arcsec) in  the
spatial  co-ordinates. The  zeroth and  first order  moment  maps were
generated  from  the  blanked  channel  images  containing  \ion{H}{i}
emission and two additional channels on both the sides.

\subsection{The VLA observations} 

The  VLA `B'  configuration observations  were carried  out  in March,
2001.  The     observational    parameters    are     summarized    in
Table~\ref{tab:obspar}. The  data were  recorded in the  1A correlator
mode   with  a   total  bandwidth   of  6.25~MHz   divided   into  128
channels.   These   observations   covered   a   velocity   range   of
$4400-5550$~km~s$^{-1}$ for  the 1667 MHz  line of OH with  a velocity
resolution of  $\sim$~9~km~s$^{-1}$.  These observations  also covered
the   1665   MHz    line   of   OH   in   the    velocity   range   of
$4050-5200$~km~s$^{-1}$.    A   large   fluctuation  in   the   system
temperature was  noticed on the  first day of observations  which were
carried out  at a centre  frequency of 1640.5~MHz.  These fluctuations
were  later  identified as  due  to  strong  signals from  a  external
interfering source operating near  the frequency of observations.  The
observations  on the  next day  were  carried out  after reducing  the
front-end bandwidth from 25~MHz to 12.5~MHz and shifting the centre of
the band to  1640.22~MHz to avoid the external  interference. The data
from the first day of observations were discarded.

The VLA observations and data  analyses were carried out following the
same  strategy  as adopted  for  the  GMRT  observations described  in
section 3.1. The  flux density of 3C~48 was  estimated to be 14.270~Jy
at  the  observing  frequency.  The  image cube  was  generated  using
``natural-weighted" continuum-free data to get maximum signal to noise
ratio. The  image cube was  box-car smoothed along the  frequency axis
using a window  of 3 channels and every  second channel was discarded.
The     resulting     image    cube     has     a    resolution     of
$5\farcs43~\times~4\farcs93~\times  27$   km~s$^{-1}$.  The  continuum
images  were made  using the  self-calibrated averaged  data  from the
line-free channels.

\section{Results} 
\subsection{Radio continuum} 

\begin{figure*}
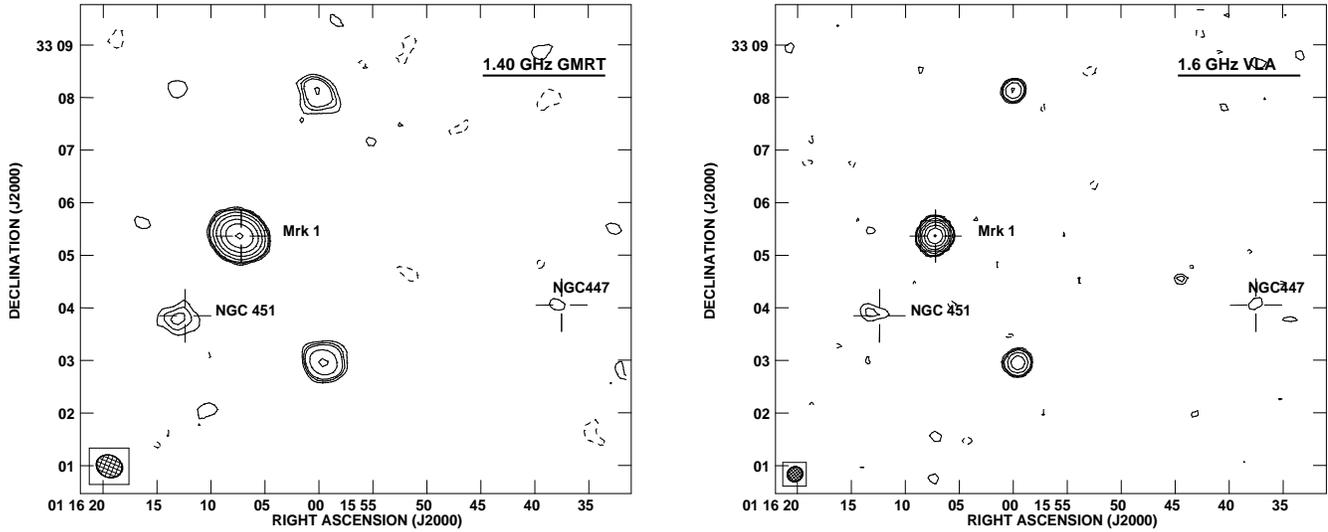

\par
\includegraphics[width=7cm,angle=-90,clip]{fig1.epsi}
\hspace*{8mm}
\includegraphics[width=7cm,angle=-90,clip]{fig2.epsi}

\caption{Radio continuum images of the group WBL~035. The image in the
left panel is at 1.4~GHz from the  GMRT and that at the right panel is
at 1.6~GHz from the VLA. The crosses mark the optical positions of the
members of the group.  The contours are drawn as -1, 1,  1.5, 2, 4, 8,
16,  32, 64, 128  in units  of 1.1~mJy~beam$^{-1}$  at 1.4~GHz  and in
units   of   0.5~mJy~beam$^{-1}$   at  1.64~GHz   respectively.    The
synthesized  beam,   shown  in  the   bottom  left  hand   corner,  is
$30\farcs75~\times~25\farcs17$, PA =  +67.8\degr in the 1.4~GHz image;
and  $17\farcs49~\times~16\farcs56$, PA =  --52.0\degr in  the 1.6~GHz
image.  The   images  are   corrected  for  respective   primary  beam
attenuations.  The rms  noise is  0.35~mJy~beam$^{-1}$ in  the 1.4~GHz
image  and 0.19~mJy~beam$^{-1}$ in  the 1.6~GHz  image. The  peak flux
densities  in  the  images  are 76.3~mJy~beam$^{-1}$  at  1.4~GHz  and
68.1~mJy~beam$^{-1}$ at 1.6~GHz.}

\label{fig:radcont} 
\end{figure*}

The  radio continuum  images  shown in  Fig.~\ref{fig:radcont} have  a
resolution   of   $30\farcs75~\times~25\farcs17$   at   1.4~GHz,   and
$17\farcs49~\times~16\farcs56$  at 1.6~GHz.   These  images were  made
using only short (u,v) spacings to enhance the sensitivity to extended
features.   These  images have  an  rms  of $0.35$~mJy~beam$^{-1}$  at
1.4~GHz, and $0.19$~mJy~beam$^{-1}$  at 1.6~GHz. Continuum emission is
detected from  both Mrk~1 and NGC~451. NGC~447  is marginally detected
($\sim4\sigma$) at both 1.4 and 1.6~GHz.  The flux density of Mrk~1 is
estimated  to   be  $76\pm4$~mJy  at  1.4~GHz,   and  $68\pm3$~mJy  at
1.6~GHz. The spectral index between  1.4 and 1.6~GHz is 0.8.  The flux
density of  NGC~451 is $3.3\pm0.5$~mJy at  1.4~GHz and $1.8\pm0.4$~mJy
at  1.6~GHz.   Mrk~1  remains  unresolved  down  to  a  resolution  of
$\sim1$~kpc.

\subsection{\ion{H}{i} emission}

\begin{figure*}
\centering
\includegraphics[width=18cm,angle=-90]{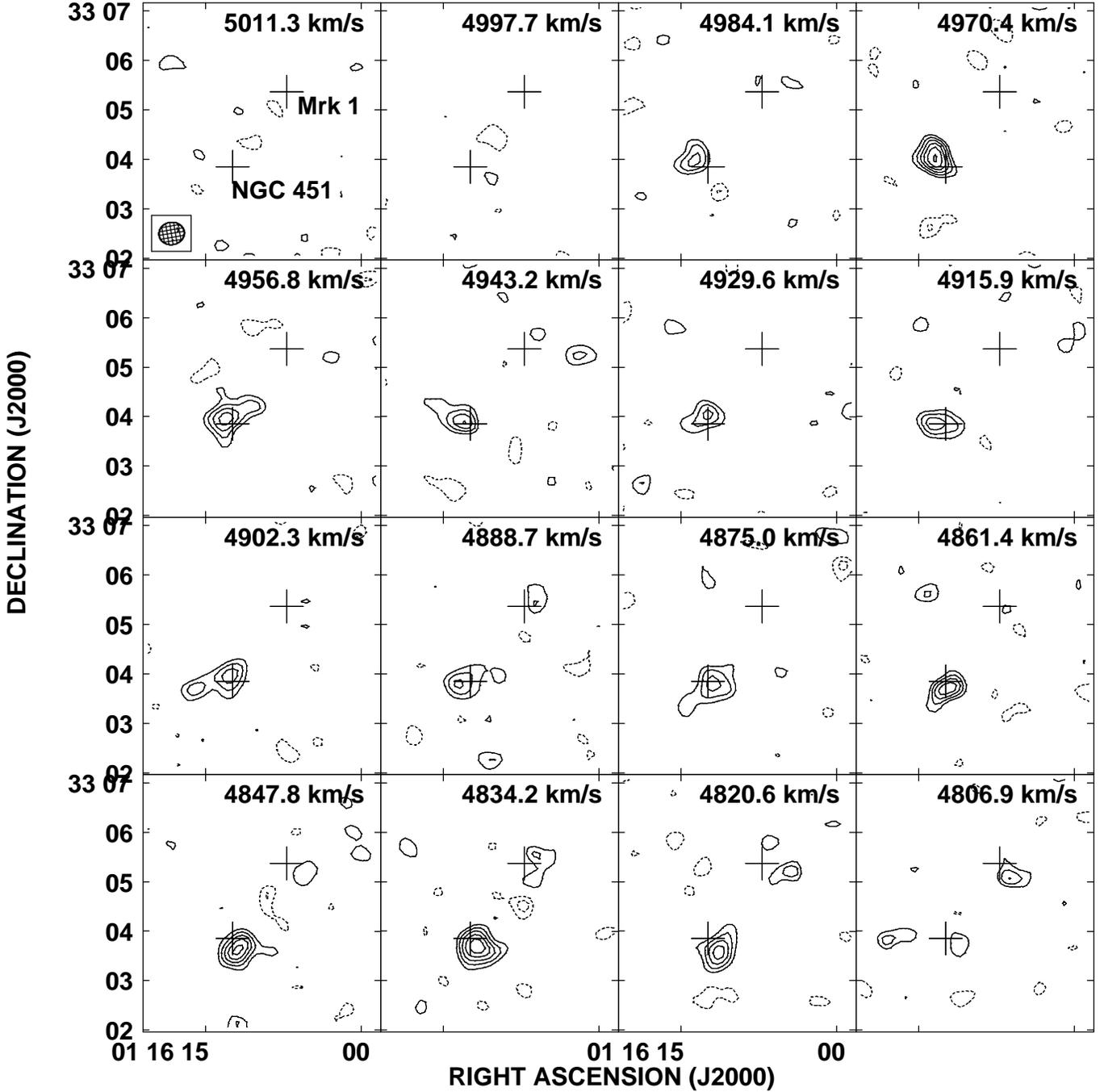}

\caption{Channel  images from GMRT  showing \ion{H}{i}  column density
contours    in     the    velocity    range     4807~km~s$^{-1}$    --
5011~km~s$^{-1}$. The crosses mark  the optical positions of Mrk~1 and
NGC~451.   Solid contours  representing column  density  of \ion{H}{i}
emission  are   drawn  at   3.6,  5.4,  7.2,   9.0,  10.8,   and  12.7
$\times$10$^{19}$~cm$^{-2}$. The negative contours (dashed curves) are
drawn at 2,  3, 4, 5, 6 mJy~beam$^{-1}$.  The  HPBW of the synthesized
beam ($30\farcs67~\times27~\farcs28$,  PA = --80.4\degr)  is indicated
at  the bottom  left  hand corner  of  the first  channel image.   The
velocity resolution in the cube is $\sim$~13.7~km~s$^{-1}$. }

\label{fig:chimage1}
\end{figure*}

\begin{figure*}
\centering
\includegraphics[width=18cm,angle=-90]{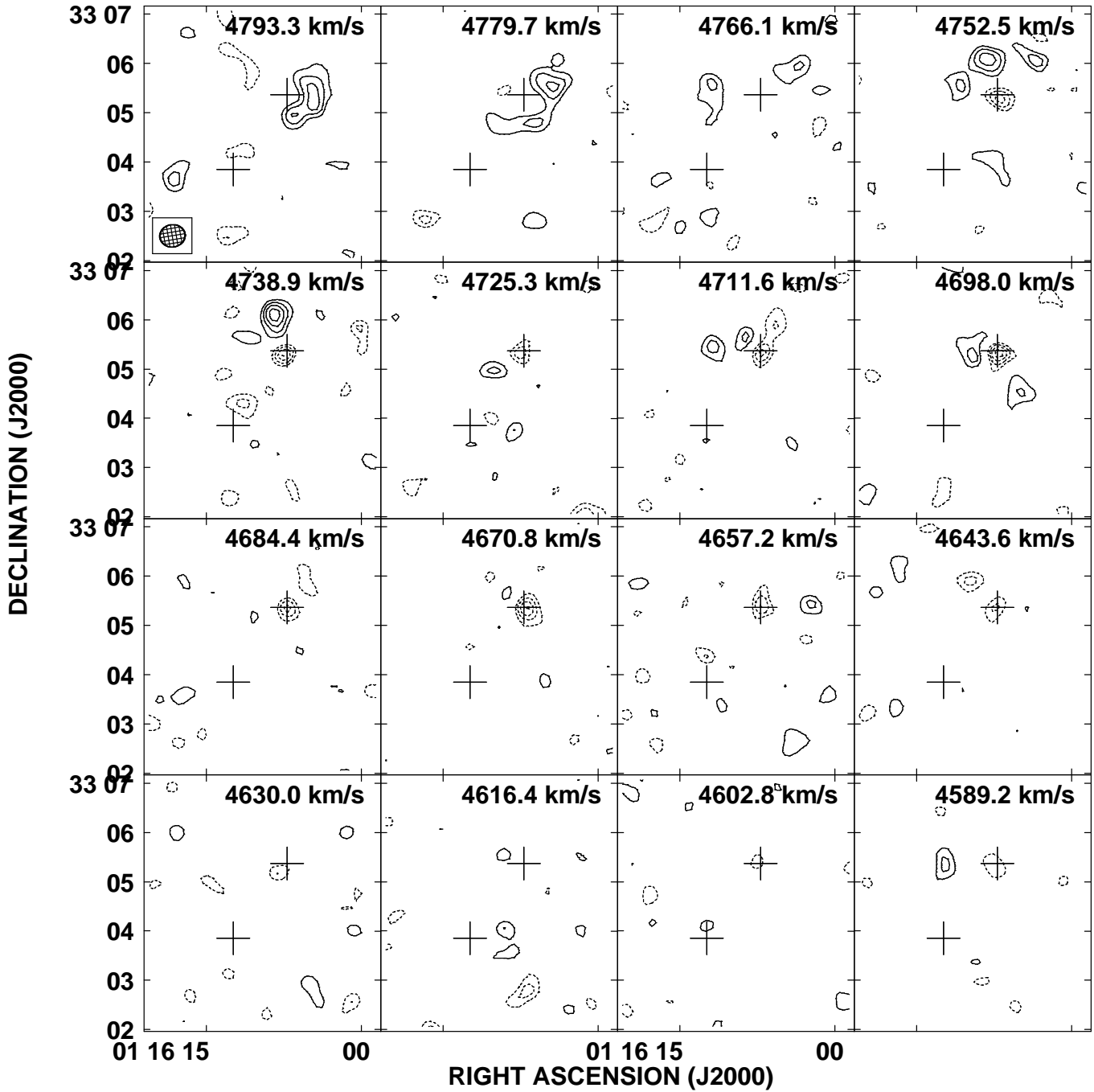}

\caption{Channel  images  showing   column  density  contours  in  the
velocity  range  4589~km~s$^{-1}$  -- 4793~km~s$^{-1}$.   The  contour
levels  are the  same as  in Fig.~\ref{fig:chimage1}.   The \ion{H}{i}
absorption is seen toward Mrk~1 as dotted contours.}

\label{fig:chimage2}
\end{figure*}

\begin{figure}
\centering
\includegraphics[height=7cm,width=9cm,angle=0]{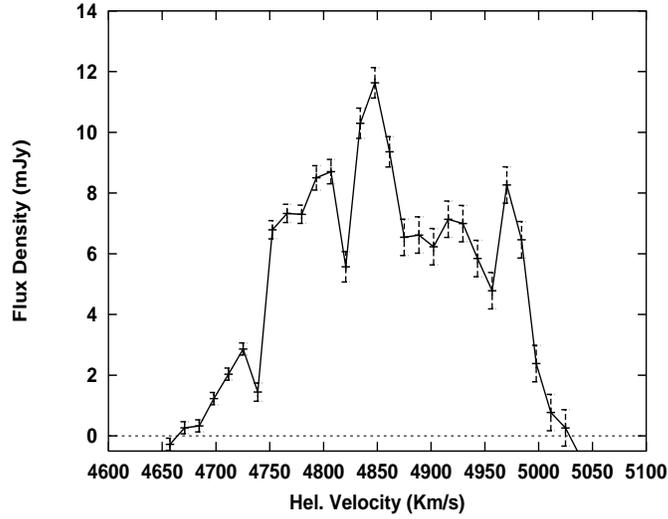}

\caption{Global \ion{H}{i} emission profile  of Mrk~1 and NGC~451 from
GMRT.  The  flux  integral  is $1.93\pm0.11$~Jy~km~s$^{-1}$  which  is
consistent with the single dish observations of \cite{hut89}.  }

\label{fig:hiems}
\end{figure}

\begin{figure}
\centering 
\includegraphics[width=8.5cm, angle=0]{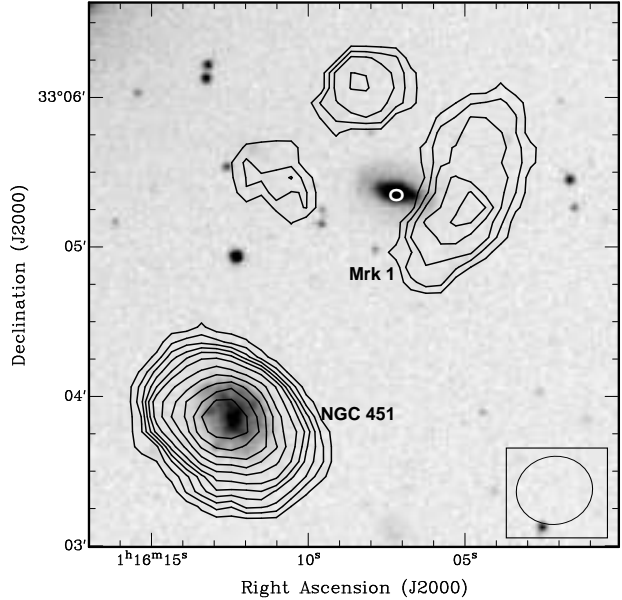} 

\caption{The  column density  contours of  the total  \ion{H}{i} image
from GMRT of  Mrk~1 (top) and NGC~451 (bottom)  overlaid upon the grey
scale optical image  from the DSS (blue). The  contour levels are 0.3,
0.8,  1.3,  1.8,  2.3,  3,  4,  5,   6,  7,  8,  and  9  in  units  of
$10^{20}$~cm$^{-2}$.  The  HPBW of the synthesized beam,  shown in the
bottom  right  hand corner,  is  $30\farcs67~\times~27\farcs28$, PA  =
--80.4\degr.  Although the \ion{H}{i}  emission is  surrounding Mrk~1,
\ion{H}{i}  absorption (marked  as  white circle)  is detected  toward
Mrk~1 indicating the  presence of cold \ion{H}{i} gas  in front of it.
}

\label{fig:mom0}
\end{figure}

\begin{figure}
\centering
\includegraphics[width=9cm, angle=0,clip]{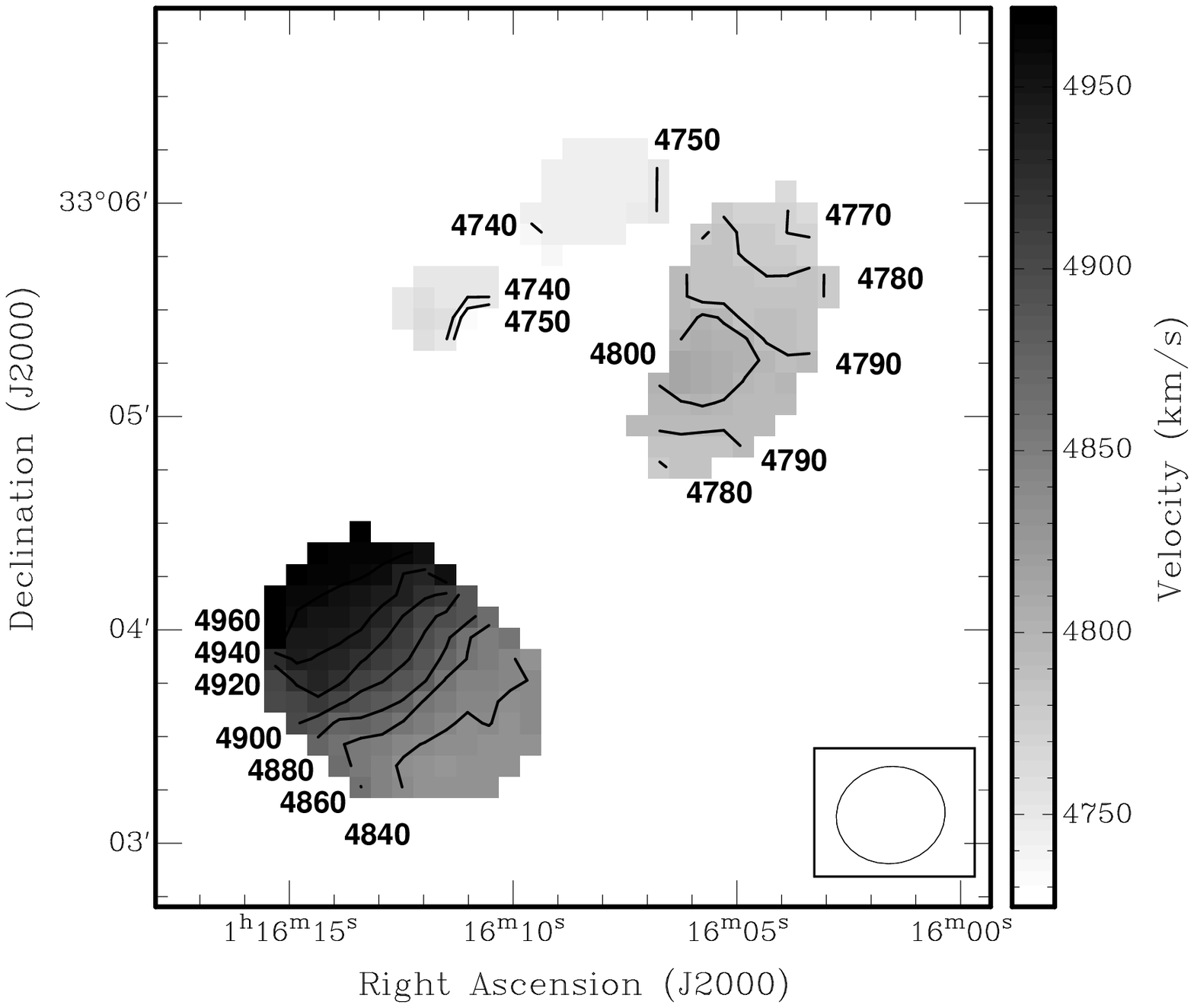}

\caption{The velocity fields of Mrk~1 and NGC~451 from GMRT are shown as contours and in grey
scale. }

\label{fig:mom1}
\end{figure}

The    \ion{H}{i}   cube    was    made   with    a   resolution    of
$30\farcs67~\times~27\farcs28~\times~13.7$~km~s$^{-1}$.   The  channel
images  have   an  rms  of   0.92~mJy~beam$^{-1}$.  The  corresponding
3$\sigma$    sensitivity    in    \ion{H}{i}   column    density    is
$5.0~\times~10^{19}$~cm$^{-2}$.   The  channel  images  of  \ion{H}{i}
emission  and  absorption are  shown  in Figs.~\ref{fig:chimage1}  and
~\ref{fig:chimage2}.      The     results     are    summarized     in
Table~\ref{tab:rad+hiem}.   \ion{H}{i}  emission  is detected  in  the
velocity range  of 4698 to 4984 km~s$^{-1}$.  \ion{H}{i} emission from
Mrk~1 appears from 4698 to  4848 km~s$^{-1}$. The flux integral ($\int
S~dV$)  of   \ion{H}{i}  emission  in  this   range  is  $0.73\pm0.05$
Jy~km~s$^{-1}$ which at  the distance of Mrk~1 corresponds  to a total
\ion{H}{i}  mass of  $8.0(\pm0.6)~\times~10^{8}$~M$_\odot$  for Mrk~1.
The  estimated  HI  mass is  a  lower  limit  due  to the  effects  of
absorption. \ion{H}{i} emission from  NGC~451 is detected from 4807 to
4984  km~s$^{-1}$  with a  total  flux  integral  of $1.20\pm0.10$  Jy
km~s$^{-1}$     corresponding    to     a    \ion{H}{i}     mass    of
$1.31(\pm0.11)~\times~10^{9}$~M$_\odot$ assuming a distance to NGC~451
of  68~Mpc.  The  summed  flux   integral  of  Mrk~1  and  NGC~451  is
$1.93(\pm0.11)$  Jy km~s$^{-1}$,  which is  consistent with  the value
obtained  by  the  single   dish  observations  of  \cite{hut89}.  See
Fig.~\ref{fig:hiems} for a global \ion{H}{i} profile.

The moment  zero map shown  in Fig.~\ref{fig:mom0} indicates  that the
\ion{H}{i}  emission from Mrk~1  is distributed  in three  clumps with
almost all  \ion{H}{i} seen outside  the optical extent of  Mrk~1. The
individual clumps having velocity  dispersions of 30 to 60 km~s$^{-1}$
are distributed over an extent of $\sim30$~kpc.  The velocity field of
Mrk~1 shown  in Fig.~\ref{fig:mom1} indicates  that there is  a smooth
rotation of \ion{H}{i} from one end to the other.

The  \ion{H}{i}  emission from  NGC~451  shown in  Fig.~\ref{fig:mom0}
looks like that of a disk galaxy with a total projected velocity width
of 170~km~s$^{-1}$.   The \ion{H}{i}  diameter of NGC~451  is $\sim$20
kpc  which  is  about twice  that  of  the  optical disk.  The  global
parameters of  NGC~451 given in  Table~\ref{tab:rad+hiem} were derived
from  a fit  to  the velocity  field  made using  a higher  resolution
($17\arcsec\times14\arcsec$) \ion{H}{i} cube which is not shown here.

\begin{table}
\caption{\bf{Radio continuum and \ion{H}{i} emission results}}
\label{tab:rad+hiem}
\begin{tabular}{lcc}
\hline
\hline
\bf{Parameter}& \bf{Mrk~1} & \bf{NGC~451}\\
\hline
S$_{1.4 GHz}^{\mathrm{(cont.)}}$  (mJy) &$76\pm4$ &$3.3\pm0.5$ \\
S$_{1.6 GHz}^{\mathrm{(cont.)}}$  (mJy) &$68\pm3$ &$1.8\pm0.4$ \\
$\delta$V$^{\mathrm{(\ion{H}{i}~emission)}}$ (km s$^{-1}$) &120 &170 \\
Systemic velocity (km s$^{-1}$) &$4780\pm13^{\dagger}$ &$4897\pm2^{\ddagger}$ \\
Maximum rot. velocity (km s$^{-1}$) &--&$140^{\ddagger}$\\\
Inclination (degree) &--&$31\pm10^{\ddagger}$\\
\ion{H}{i} extent (kpc) &$\sim30$ &$\sim20$\\
$\int S^{\mathrm{(\ion{H}{i}~emission)}}$~dV (Jy km s$^{-1}$)&$0.73\pm0.05$ &$1.2\pm0.1$ \\
\ion{H}{i} mass (10$^{8}$ M$_\odot$)&$8.0\pm0.6$ &$13\pm1.0$ \\
\hline
\multicolumn{3}{p{3in}}{$\dagger$: Estimated from the mean velocity of \ion{H}{i} emission}\\
\multicolumn{3}{p{3in}}{$\ddagger$: Estimated from a fit of curves of constant velocities to 
the \ion{H}{i} velocity field}
\end{tabular}
\end{table}

\subsection{\ion{H}{i} absorption}

\begin{figure}
\centering
\includegraphics[height=7cm,width=9cm,angle=0]{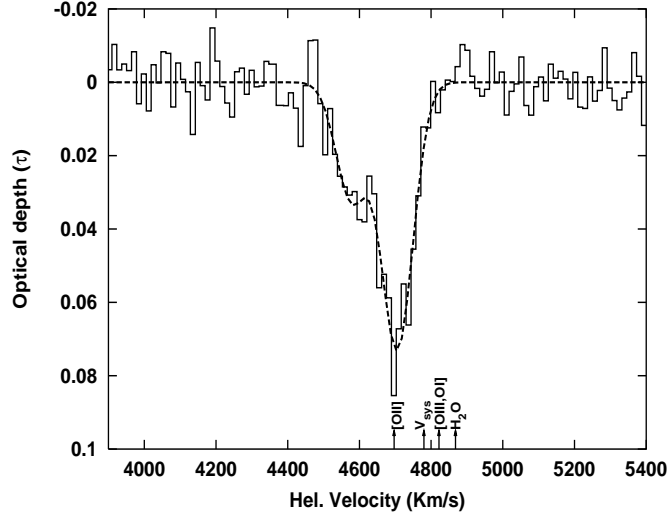}

\caption{The GMRT spectrum showing \ion{H}{i} absorption in Mrk~1. The
dotted  curve is  the Gaussian  fit  to the  absorption spectrum.  The
fitted  parameters  are given  in  Table~\ref{tab:spec}. The  vertical
lines along the  velocity axis mark the positions  of several velocity
systems as indicated.}

\label{fig:hispec}
\end{figure}

\begin{figure}
\centering
\includegraphics[height=7cm,width=9cm,angle=0]{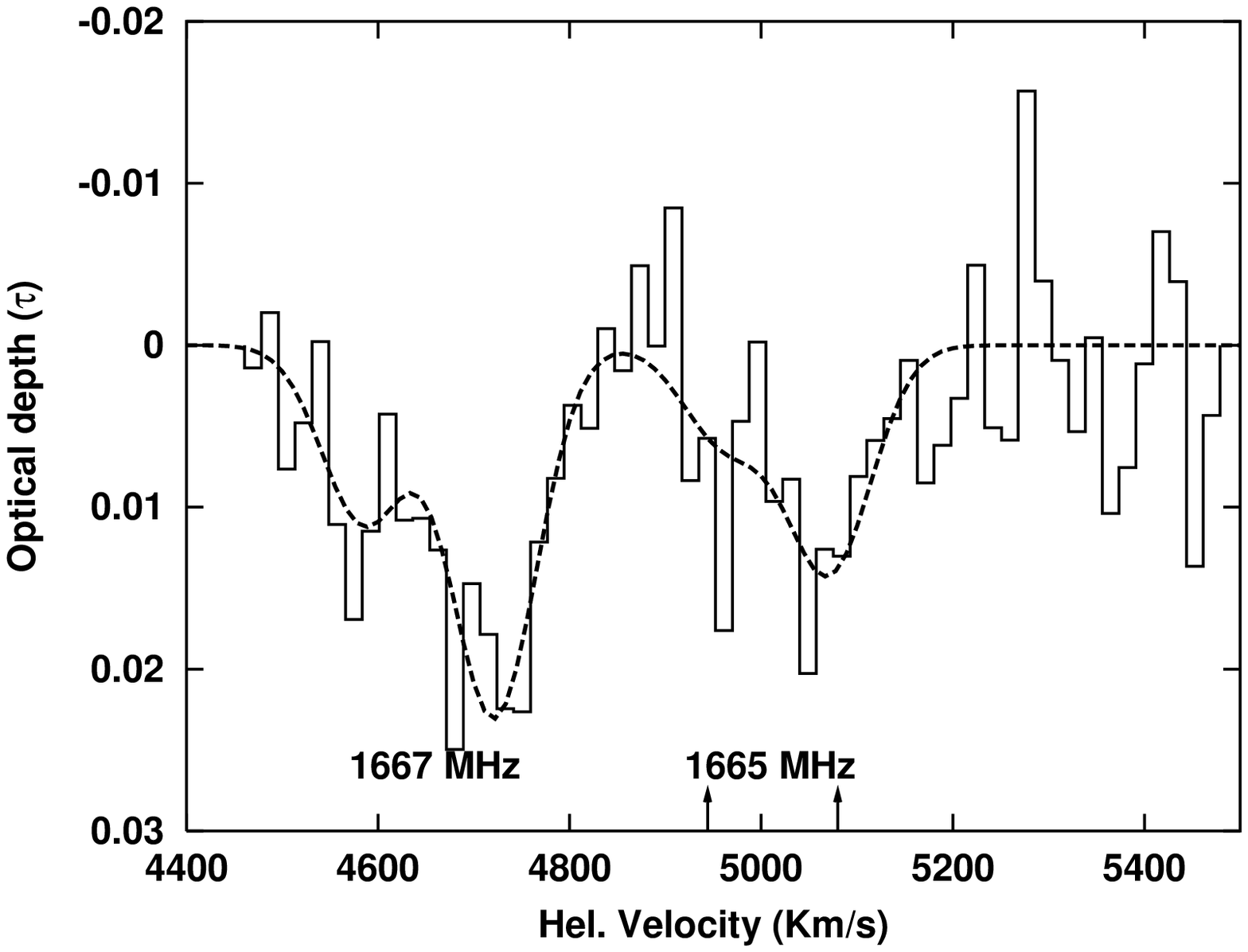}

\caption{The  VLA  spectrum  showing  OH absorption  from  Mrk~1.  The
velocity axis corresponds to the 1667~MHz line. The dotted curve shows
the  model spectrum  of the  1665 and  1667~MHz OH  lines.  The fitted
parameters  are  given   in  Table~\ref{tab:spec}.  In  this  velocity
definition, the 1665~MHz line will appear at +360 km~s$^{-1}$ from the
1667~MHz  line. Two  vertical  lines  in the  velocity  axis mark  the
expected  positions  of   1665~MHz  absorption  corresponding  to  the
1667~MHz absorption detected near 4721 and 4585 km~s$^{-1}$.}

\label{fig:ohspec}
\end{figure}
 
The  channel   images  shown  in   Fig.~\ref{fig:chimage2}  also  show
\ion{H}{i} absorption  from Mrk~1 (dotted contours at  the location of
Mrk~1) in the velocity range 4589~km~s$^{-1}$ to 4752~km~s$^{-1}$. The
\ion{H}{i}  absorption  spectrum  shown  in  Fig.~\ref{fig:hispec}  is
extracted at the  radio position of Mrk~1 from  a \ion{H}{i} cube made
with                  a                  resolution                 of
$6\farcs23~\times~4\farcs39~\times~13.7$~km~s$^{-1}$. The  rms in this
cube   was   0.6~mJy~beam$^{-1}$.   The   spectrum   shows   a   broad
multi-component absorption in  between the velocities 4500 km~s$^{-1}$
and  4800 km~s$^{-1}$.  Two  Gaussian components  were  fitted to  the
\ion{H}{i} absorption  profile.  The  resulting parameters of  the fit
are given in Table~\ref{tab:spec}. The  peak optical depths of the two
components  are $0.073\pm0.004$  and $0.032\pm0.004$  respectively and
the  velocity  dispersions  are  $\sim$~44~km~s$^{-1}$  for  both  the
components. The  column density of  \ion{H}{i} is estimated  using the
relation                    $N_\mathrm{\ion{H}{i}}                   =
1.82~\times~10^{18}~\times~(T_\mathrm{spin}/f)~\int   \tau~\mathrm{d}v$
~cm$^{-2}$;  where  $T_\mathrm{spin}$   is  the  spin  temperature  of
\ion{H}{i} in kelvin, $f$ is  the covering fraction of \ion{H}{i} gas,
$\int\tau~\mathrm{d}v$ is the velocity  integrated optical depth in km
s$^{-1}$. We assume  $f$ to be unity. $T_\mathrm{spin}$  is an unknown
quantity and we adopt a value  of 100~K, typical of cold clouds in our
Galaxy.     The    \ion{H}{i}     column     densities    are     then
$1.5(\pm0.2)~\times~10^{21}$~cm~$^{-2}$                             and
$6.0(\pm1.5)~\times~10^{20}$~cm~$^{-2}$   for   the   two   components
respectively.

\begin{table}[h]
\caption{\bf{\ion{H}{i} and OH absorption results}}
\label{tab:spec}
\begin{tabular}{lccc}
\hline
\hline
\bf{Parameter}& \bf{\ion{H}{i}} & \bf{OH(1667)} & \bf{OH(1665)}\\
\hline
\bf{$\tau_{1}$} &$0.073\pm0.004$ &$0.023\pm0.003$ &$0.014\pm0.004$ \\
\bf{$v_{1}$}~(km~s$^{-1}$)&$4705\pm5$ &$4721\pm6$ &$4710\pm19$ \\
\bf{$\sigma_{v_{1}}$}~(km~s$^{-1}$) &$44.2\pm4.4$ &$44.2$ &$44.2$ \\
\bf{$\tau_{2}$} &$0.032\pm0.004$ &$0.011\pm0.003$ & $0.006\pm0.004$\\
\bf{$v_{2}$}~(km~s$^{-1}$) &$4579\pm10$ &$4585\pm12$ & $4601\pm45$ \\
\bf{$\sigma_{v_{2}}$}~(km~s$^{-1}$) &$43.4\pm9.1$ &43.4 & $43.4$\\
\hline
\end{tabular}
\end{table}

\subsection{OH absorption}

The OH  spectrum shown in  Fig.~\ref{fig:ohspec} was extracted  at the
radio position  of Mrk~1 from the  image cube made using  the VLA data
with                  a                  resolution                 of
$5\farcs43~\times~4\farcs93~\times~27$~km~s$^{-1}$.  The  cube has  an
rms of 0.5~mJy~beam$^{-1}$.

The velocity axis of Fig.~\ref{fig:ohspec} corresponds to the 1667~MHz
OH line.  In this  velocity system, the  1665~MHz line will  appear at
+360~km~s$^{-1}$  from  the  1667~MHz  line.  Since  the  spectrum  of
Fig.~\ref{fig:ohspec} does  not have enough baseline  for the 1667~MHz
line and not  enough signal to noise ratio for  both the 1665~MHz line
to  get  reliable estimates  for  velocity  dispersions of  individual
components, only peak optical depths and center velocities were fitted
while the  velocity dispersions  were fixed at  those values  found in
fitting the  \ion{H}{i} absorption profile. This is  a reasonably good
assumption since both the 1665 and 1667 MHz profiles are qualitatively
similar to  the \ion{H}{i} absorption  profile. This procedure  gave a
reasonably  good, though  not unique,  fit  to the  OH spectrum.   The
fitted parameters are given in Table~\ref{tab:spec}.

The  peak  optical  depths of  the  two  components  of the  1667  MHz
absorption are $0.023\pm0.003$  and $0.011\pm0.003$ respectively.  The
column density of OH is  estimated using the relation $N_\mathrm{OH} =
2.35~\times~10^{14}~\times~(T_\mathrm{ex}/f)~\int
\tau_\mathrm{1667}~\mathrm{d}v$  ~cm$^{-2}$; where  $T_\mathrm{ex}$ is
excitation   temperature   which   is   assumed   to   be   10~K   and
$\int~\tau_\mathrm{1667}~\mathrm{d}v$   is  the   velocity  integrated
optical depth  of the  1667~MHz line in  units of km~s$^{-1}$.  The OH
column        densities        are        estimated       to        be
$6.0(\pm1.0)~\times~10^{15}$~cm$^{-2}$                              and
$2.9(\pm0.9)~\times~10^{15}$~cm$^{-2}$.  The peak optical  depth ratio
$\tau_\mathrm{1667}/\tau_\mathrm{1665}$  of the stronger  OH component
is $1.6\pm0.5$,  indicating that this  component is excited  under LTE
conditions -- the ratio is predicted  to be in between 1.0 and 1.8 for
LTE  excitations.   This  ratio   for  the  weaker   component,  viz.,
$1.8\pm1.3$ indicates that this is also most likely thermally excited.

\section{Discussion}

\subsection{Interaction of Mrk~1 with NGC~451}

The  disturbed  \ion{H}{i}  morphology of  Mrk~1  (Fig~\ref{fig:mom0})
indicates  a  gravitational  interaction  possibly  with  the  nearest
companion  NGC~451. We  explore this  possibility using  the  two body
interaction  described  in   \cite{bin87}.  The  dynamical  masses  of
galaxies are  estimated using  rotation curves.  Since  the \ion{H}{i}
morphology  of Mrk~1 is  disturbed, it  was not  possible to  obtain a
reliable  \ion{H}{i} rotation  curve. We  used the  H$\alpha$ rotation
curve  of  Mrk~1 (Keel  1996).   The  dynamical  mass of  NGC~451  was
estimated using  the \ion{H}{i} rotation curve.   The dynamical masses
of   Mrk~1  and   NGC~451  are   $3.4~\times~10^{10}$~M$_{\odot}$  and
$4.5~\times~10^{10}$~M$_{\odot}$    respectively.    The   interaction
parameters are listed in Table~\ref{tab:int}.

\begin{table}
\centering
\caption{\bf{Interaction properties of the Mrk~1--NGC~451 system}}
\begin{tabular}{lc}
\hline
\hline
\bf{Parameter} & \bf{Value} \\
\hline
Projected velocity difference (km~s$^{-1}$)& 117 \\ 
Projected  separation (kpc)  &38 \\ 
Total dynamical mass (M$_{\odot}$) & $\sim10^{11}$  \\
Tidal radius (Mrk~1) (kpc) &24 \\
Tidal radius (NGC~451) (kpc) &29\\
Impact parameter  & $\sim0.1$ \\
Dynamical friction time (Gyr)& $\sim$ 0.2  \\
Orbital time (Gyr) & $\sim$ 2.0  \\
\hline
\label{tab:int}
\end{tabular}
\end{table}

The  projected  velocity  difference  between  Mrk~1  and  NGC~451  of
$\sim$~117~km~s~$^{-1}$  indicates a  minimum dynamical  mass  of this
pair to be $\sim~10^{11}$~M$_{\odot}$. This value of dynamical mass is
in close  agreement with  the dynamical masses  of Mrk~1  and NGC~451,
indicating  that  Mrk~1  and  NGC~451  are  most  likely  in  a  bound
system. Tidal radii  (cf. eq.  7-84, Binney \&  Tremaine 1987) for the
given masses of  Mrk~1 and NGC~451 indicate that  the outer regions of
the \ion{H}{i} disk  of Mrk~1 can be perturbed  easily.  The dynamical
friction time scale (cf. eq.  7-26 and 7-13b, Binney \& Tremaine 1987)
of $\sim0.2$~Gyr  for this system  is much smaller than  their orbital
time  scale of  $\sim$~2~Gyr.  This  implies that  the  interaction is
bound and will lead to a merger within a small fraction of the orbital
time period of the two galaxies.

\subsection{Comparison of \ion{H}{i} and OH velocities with other 
velocity systems}

\noindent$Comparison~with~optical~line~velocities-$      From      the
comparison  of the \ion{H}{i}  and OH  absorption velocities  of Mrk~1
with the  optical line velocities listed  in Table~\ref{tab:mrk01}, it
appears that the  higher velocity components of the  \ion{H}{i} and OH
absorption  are consistent  with  the [\ion{O}{ii}]$\lambda3727$  line
velocity,  but  are  blueshifted  by $\sim$~100~km~s$^{-1}$  from  the
[\ion{O}{iii}]$\lambda5007$,    [\ion{O}{i}]$\lambda6300$    and   the
systemic  velocity.    We  explain   this  discrepancy  in   terms  of
co-existence  of  photo-ionized  and   shock  ionized  gas  in  active
galaxies.  The [\ion{O}{iii}]$\lambda5007$  line is  primarily  due to
excitation  from a hard  continuum, and  therefore, should  be arising
close to the nucleus. The [\ion{O}{ii}]$\lambda3727$ line intensity is
enhanced in shock ionized  regions (\cite{dop95}). Most often, optical
line profiles are asymmetric and  only peak line velocities are quoted
without  fitting a  line profile.  Mrk~1 is  known to  be such  a case
(Bergeron  \& Durret 1987,  \cite{dick76}).  Such  an analysis  of the
optical  spectrum may bias  the line  velocities of  different species
toward different regions, e.g., the peak of the [\ion{O}{ii}] line may
indicate a region which is shock ionized while the [\ion{O}{iii}] line
velocity may indicate gas which  is close to the nucleus. We speculate
that  the higher velocity  \ion{H}{i} and  OH absorption  component in
Mrk~1  arises in  a  region which  is  pushed forward  due to  shocks,
thereby  giving   an  apparent   blueshift  to  \ion{H}{i},   OH,  and
[\ion{O}{ii}] lines.  The fact that the [\ion{O}{i}]  line velocity is
close to  the [\ion{O}{iii}] line velocity, and  hence associated with
photo-ionized regions,  is not surprising since  the [\ion{O}{i}] line
intensity is suppressed in the shock excited regions (\cite{dop95}).

\noindent$Comparison~with~H_{2}O~megamaser~and~CO~emission-$ The water
megamasers  are seen  from  Mrk~1 at  a  velocity of  4868~km~s$^{-1}$
(\cite{bra94}).    Since    these    masers    are    redshifted    by
$\sim$~90~km~s$^{-1}$ from  the systemic velocity of  the galaxy, they
are most likely the  high velocity ``satellite" features commonly seen
in water megamaser galaxies (\cite{bra97}) and thought to originate in
the accretion disks near the nuclei (\cite{neu94}). Since the observed
\ion{H}{i} and OH absorption  velocities in Mrk~1 are blueshifted from
both the  water megamaser velocity  and from the systemic  velocity of
the  galaxy, the absorption  in the  present case  is most  likely not
related either to the gas in  the accretion disk or to the torus close
to the  nucleus. The mean  velocity of the  CO emission from  Mrk~1 is
4850   km   s$^{-1}$    (\cite{vil98}),   which   is   redshifted   by
$\sim$~150~km~s$^{-1}$ from  \ion{H}{i} and OH  absorption velocities,
implying that the gas traced via \ion{H}{i} and OH absorption in Mrk~1
is also not related to the molecular gas traced by CO emission.

\subsection{Kinematics and composition of the absorbing gas}

\begin{table}
\centering
\caption{\bf{Properties of the absorbing gas}}
\begin{tabular}{lc}
\hline
\hline
\bf{Parameter} & \bf{Value} \\
\hline
$N_\mathrm{1}$(\ion{H}{i})$^\dagger$ (cm$^{-2}$) &$1.5(\pm0.2)~\times~10^{21}$ \\
$N_\mathrm{2}$(\ion{H}{i})$^\dagger$ (cm$^{-2}$) &$6.0(\pm1.5)~\times~10^{20}$ \\
$N_\mathrm{1}$(OH)$^\ddagger$ (cm$^{-2}$) &$6.0(\pm1.0)~\times~10^{15}$ \\
$N_\mathrm{2}$(OH)$^\ddagger$ (cm$^{-2}$) &$2.8(\pm1.0)~\times~10^{15}$ \\
$N_\mathrm{total}$(\ion{H}{i})  (cm$^{-2}$) &$2.1(\pm0.6)~\times~10^{21}$ \\
$N_\mathrm{total}$(OH)  (cm$^{-2}$) &$8.8(\pm3.4)~\times~10^{15}$ \\
$\tau_\mathrm{\ion{H}{i}}/\tau_\mathrm{OH,~1667}$ (1) &$3.2\pm0.4$ \\
$\tau_\mathrm{\ion{H}{i}}/\tau_\mathrm{OH,~1667}$ (2) &$2.9\pm0.9$ \\ 
$\tau_\mathrm{OH,~1667}/\tau_\mathrm{OH,~1665}$ (1) & $1.6\pm0.5$\\
$\tau_\mathrm{OH,~1667}/\tau_\mathrm{OH,~1665}$ (2) & $1.8\pm1.3$\\
$N_\mathrm{total}$ (H$_\mathrm{2}$)$^\dagger$$^\dagger$  (cm$^{-2}$)  &$\sim10^{23}$ \\
$N_\mathrm{total}$(OH)/$N_\mathrm{total}$(\ion{H}{i}) &$\sim4.3~\times~10^{-6}$ \\
A$_\mathrm{v}$  (mag)& 9--90 \\
$N_\mathrm{H}$  cm$^{-2}$ &$\sim1.1~\times~10^{23}$ \\
$\tau_\mathrm{photoelectric}^\mathrm{1keV}$ & $\sim$~30 \\
\hline
\\
\multicolumn{2}{p{3in}}{$\dagger$  : Assuming $T_\mathrm{spin}$=100~K;
$\ddagger$   :  Assuming   $T_\mathrm{ex}$=10~K;   $\dagger\dagger$  :
Assuming OH/H$_{2}$=10$^{-7}$}
\label{tab:absparam}
\end{tabular}
\end{table}

The general properties of the gas seen in absorption are summarized in
Table~\ref{tab:absparam}.    The  total  column   density  of   OH  is
comparable to that observed in other active galaxies (e.g., Schmelz et
al.   1986, Baan  et  al.  1985,  1992).   Both components  of the  OH
transitions  appear to  be thermally  excited as  their  optical depth
ratios ($\tau_{1667}/\tau_{1665}$) are between 1.0 and 1.8; the values
predicted for excitations in  LTE conditions. The optical depth ratios
$\tau_\mathrm{\ion{H}{i}}/\tau_\mathrm{OH}$ for both the components of
the  absorbing gas  are  $\sim$~3. This  ratio  has been  found to  be
varying  from  as  low as  5  to  more  than  400 in  Galactic  clouds
(\cite{dic81}). The smaller values  correspond to the molecular clouds
while larger values correspond to  the diffuse clouds. It is therefore
suggested that the \ion{H}{i} and  OH absorption, in the present case,
are associated with dense molecular clouds.

The  observed  velocity  dispersion  ($\sigma$)  of  the  1667~MHz  OH
absorption, viz.,  44 km~s$^{-1}$ is higher than  the typical velocity
dispersions ($\sigma$  = 3--7  km~s$^{-1}$) in giant  molecular clouds
(GMCs) of the Galactic  disk.  However, several high dispersion clouds
($\sigma\sim40$ km~s$^{-1}$) have been detected in 18 cm OH absorption
within  a kpc  of the  Galactic centre  (\cite{boy94}).   The simplest
explanation  for such  a high  velocity dispersion  could be  a chance
alignment  of  several  normal  GMCs  along the  line  of  sight,  but
\cite{kum97}  have shown that  the probability  of such  alignments is
small. Alternatively,  if the velocity  dispersion is due to  a single
gravitationally bound  system in virial equilibrium, the  mass of such
an  object   (assuming  a  size  of  50~pc)   could  be  $\sim~10^{7}$
M$_{\odot}$.  Cloud--cloud  collisions (\cite{kle94b}) and interaction
of  shock with  ISM  (\cite{kle94a})  are also  known  to enhance  the
internal velocity dispersions of molecular clouds.

The  OH  column  density  is   known  to  correlate  with  the  visual
extinction,  A$_\mathrm{v}$,   of  molecular  clouds   in  our  Galaxy
(\cite{mag88}). Magnani et  al. (1988) found that N(OH)/A$_\mathrm{v}$
ratios are in the range of $10^{14}-10^{15}$ cm$^{-2}$~mag$^{-1}$. For
the  OH column  density  toward  Mrk~1, these  ratios  indicate 9  $<$
A$_\mathrm{v}<$ 90 toward  the line of sight of  Mrk~1. In comparison,
Veilleux et al. (1997),  based on some infrared measurements, obtained
a  lower  limit on  A$_\mathrm{v}$  to  be  26 consistent  with  above
predictions.

Using OH/H$_{2}=10^{-7}$ (\cite{lis96}), the implied column density of
H$_{2}$ is $\sim10^{23}$ cm$^{-2}$. Using values of the photo-electric
absorption cross sections from \cite{mor83} for a gas having the solar
abundance, a total hydrogen  column density of $\sim10^{23}$ cm$^{-2}$
indicates that the optical depth for X-ray absorption at 1 keV will be
$\sim$~30. Such a  high value of the optical  depth will absorb almost
all soft X-radiation from the  nucleus of Mrk~1.  Consistent with this
prediction, Mrk~1  has not been detected  as a X-ray source  down to a
sensitivity of $\sim$ 10$^{41}$ erg~s$^{-1}$ (\cite{fab92}).

\section{Conclusions}

We have  presented the observations  of the Seyfert~2 galaxy  Mrk~1 in
the \ion{H}{i}  21 cm line  using the  GMRT and in  the OH 18  cm line
using the VLA.  Unlike the optical morphology, the \ion{H}{i} emission
morphology of Mrk~1  indicates that this galaxy is  disturbed which we
interpret  as due  to  tidal interactions  with  the nearby  companion
NGC~451.   We also showed  based on  the dynamical  study of  Mrk~1 --
NGC~451  system that  the interaction  is  bound leading  to a  merger
within  a  small fraction  of  their  orbital  time period.   This  is
consistent with  the hypothesis that the bound  interactions should be
more   efficient  in  triggering   nuclear  activities   than  unbound
interactions.  The  \ion{H}{i} and  OH absorption detected  toward the
nucleus of  Mrk~1 indicates  an outflow of  both atomic  and molecular
gas.   The  column  densities   of  the  detected  \ion{H}{i}  and  OH
absorption indicate that the line of sight toward the nucleus of Mrk~1
is  rich  in both  atomic  and molecular  gas.   The  gas detected  in
absorption is kinematically different than that traced via CO emission
and  water megamaser  emission from  Mrk~1.  We  found  evidences that
shocks  (presumably   due  to  nuclear  activities)   can  affect  the
kinematics of gas  near the nucleus. The \ion{H}{i}  and OH absorption
being    blueshifted   from    the   systemic    velocity    and   the
[\ion{O}{iii}]$\lambda5007$   velocity  while   consistent   with  the
[\ion{O}{ii}]$\lambda3727$  velocity  is understood  in  terms of  the
shock   ionization  of   gas  (which   predicts  enhancement   of  the
[{\ion{O}{ii}] line intensity)  and an outflow of ISM  in front of the
shock. Based  on the optical depth  ratios and the line  widths of the
\ion{H}{i}  and OH  absorption, we  speculate that  the  absorption is
arising in turbulent molecular clouds  of similar types as those found
near the Galactic centre.  These observations also imply that the line
of sight toward the nucleus of Mrk~1 is heavily obscured.
 
\begin{acknowledgements} 
We  thank  the  staff  of   the  GMRT  that  made  these  observations
possible. GMRT is run by the National Centre for Radio Astrophysics of
the  Tata  Institute  of  Fundamental  Research.  The  National  Radio
Astronomy Observatory is a facility of the National Science Foundation
operated under cooperative  agreement by Associated Universities, Inc.
This  research has made  use of  the NASA/IPAC  Extragalactic Database
(NED) which  is operated by the Jet  Propulsion Laboratory, California
Institute of Technology, under  contract with the National Aeronautics
and  Space  Administration.  This  research  has made  use  of  NASA's
Astrophysics  Data  System.   We  thank Dipankar  Bhattacharya  for  a
reading of  the paper  and useful comments.  We thank the  referee for
thoughtful comments.

\end{acknowledgements}

\end{document}